\documentclass[prb,twocolumn,showpacs,preprintnumbers,amsmath,floatfix,superscriptaddress]{revtex4}
 \pdfoutput=1
\usepackage{graphicx}
\usepackage{dcolumn}
\usepackage{subfigure}
\usepackage{wrapfig}
\usepackage{cancel}
\usepackage{color}

\newcommand{\BSCCO}{{Bi$_2$Sr$_2$CaCu$_2$O$_8$}}
\newcommand{\dopedBSCCO}{{Bi$_2$Sr$_2$CaCu$_2$O$_{8+\delta}$}}

\begin{document}

\title{Modulation of pairing interaction in Bi$_2$Sr$_2$CaCu$_2$O$_{8+\delta}$
by an O dopant: a density functional theory study}

\author{Kateryna Foyevtsova}
\affiliation{Institut f\"ur Theoretische Physik, Goethe-Universit\"at Frankfurt, 60438 Frankfurt am Main, Germany}
\author{H. C. Kandpal}
\affiliation{IFW Dresden, P.O. Box 270016, D-01171 Dresden, Germany}
\author{Harald O. Jeschke}
\affiliation{Institut f\"ur Theoretische Physik, Goethe-Universit\"at Frankfurt, 60438 Frankfurt am Main, Germany}
\author{S. Graser}
\affiliation{Zentrum f\"ur Elektronische Korrelationen und Magnetismus, Institut f\"ur Physik, Universit\"at Augsburg, 86135 Augsburg, Germany}
\author{H.-P. Cheng}
\affiliation{University of Florida, Gainesville, Florida 32611, USA}
\author{Roser Valent{\'\i}}
\affiliation{Institut f\"ur Theoretische Physik, Goethe-Universit\"at Frankfurt, 60438 Frankfurt am Main, Germany}
\author{P. J. Hirschfeld}
\affiliation{University of Florida, Gainesville, Florida 32611, USA}

\date{\today}

\begin{abstract}
  Scanning tunneling spectroscopy measurements on the high temperature
  superconductor Bi$_2$Sr$_2$CaCu$_2$O$_{8+\delta}$ have reported an
  enhanced spectral gap in the neighborhood of O dopant atoms. We
  calculate, within density functional theory (DFT), the change in
  electronic structure due to such a dopant.  We then construct and
  discuss the validity of several tight binding (TB) fits to the DFT
  bands with and without an O dopant. With the doping-modulated TB
  parameters, we finally evaluate the spin susceptibility and pairing
  interaction within spin fluctuation theory.  The $d$-wave pairing
  eigenvalues are enhanced above the pure system without O dopant,
  supporting the picture of enhanced local pairing around such a
  defect.
\end{abstract}

\maketitle

\section{ Introduction}

With the observation in scanning tunneling spectroscopy (STS)
measurements on high-$T_c$ Bi-based cuprate
superconductors~\cite{STM_exp} that the oxygen dopant position and the
size of the superconducting gap are correlated, it has become evident
that dopant atoms may influence superconductivity beyond their roles
as sources of mobile charge and scattering centers.  Subsequently,
Nunner {\it et al.}  \cite{Nunner} showed that many STS observations
can be explained if one assumes that dopants locally enhance the
pairing interaction. It is clearly desirable to further justify the
proposal of Ref.~\onlinecite{Nunner} by identifying the microscopic
mechanism responsible for the local pairing interaction enhancement.
Such a step would fulfill a longstanding goal of allowing the
systematic study of the pairing interaction itself by the measurement
and modelling of the effects which modulate it.

If we accept the dominance of a magnetically driven pairing mechanism
in the cuprates, the pairing enhancement is a result of the local
increase in the spin fluctuation exchange interaction in the vicinity
of a dopant, which can occur due to the dopant-induced local
structural modifications. Whether the structural and the corresponding
local electronic structure modifications can indeed enhance the local
superexchange coupling has been a subject of several recent
studies~\cite{Maska,Johnston,perturbation_PRB2008}. The problem is
typically treated by calculating the local exchange coupling constants
of the $t$-$J$ model, the large-$U$ limit of the Hubbard model, with
the assumption that the local parameters of the latter (the on-site
energy $\mu$ and the hopping integrals $t$) are modified due to the
presence of a dopant.  In Ref.~\onlinecite{perturbation_PRB2008}, by
means of the Rayleigh-Schr\"odinger perturbation expansion it was
shown for the three-band Hubbard model that the exchange coupling is
enhanced only in a certain region of the Hubbard model parameters
phase diagram.  Unfortunately, recent numerical estimates of the
parameter values of the Hubbard model with an impurity based on
electrostatic calculations \cite{Johnston} place the dopant-induced
variation in the region of the phase diagram where exchange gets
suppressed.  Though discouraging at first glance, it is clear that
this general approach contains many oversimplifications of the true
electronic structure.  In addition, the full pairing interaction,
while related to the exchange coupling $J$, can be influenced also by
dynamical spin fluctuation processes.\cite{Maieretal08} We have
therefore been stimulated to investigate further the dopant-induced
effects on the local pairing with the goal of improving the model by a
refinement of the approximations.  For instance, besides considering
the variation of the atomic on-site energies (which was assumed in
Ref.~\onlinecite{Maska}~and~\onlinecite{perturbation_PRB2008} to be
the only effect due to a dopant) one can allow for the variation of
hopping integrals near the dopant as
well. 
One can furthermore go beyond the electrostatic considerations in
calculating the inhomogeneous Hubbard model parameters in order to
place more accurately the exchange coupling variations in the
parameter phase diagram by employing {\it ab initio} calculations.
Finally, one can directly calculate the dynamical pairing interaction
within weak coupling spin fluctuation theory. These are the aims of
the present work.

In this paper, we perform DFT calculations for {\dopedBSCCO} and
explore the possibility of extracting out of the electronic structure
reliable effective Cu-Cu hopping integrals $t$ and on-site energies
$\mu$ as a function of oxygen doping. In general, DFT calculations of
the electronic structure near the Fermi surface (FS) of cuprate
materials are not considered reliable due to strong correlation
effects, but it has been argued that {\it changes} in electronic
structure induced by high energy impurity states are much less
sensitive. These effective parameters are further considered in the
Hubbard model calculations of the spin susceptibility and
superconducting gap function, and key changes in these functions are
observed which support the importance of the effect of the oxygen
dopants on the pair correlations in these materials.

The evaluation of the hopping integrals and on-site energies of the
Hubbard model is performed by mapping the eigenvalues of the
corresponding non-interacting tight-binding Hamiltonian to the valence
bands of {\dopedBSCCO} as a function of oxygen doping.  Although
{\dopedBSCCO} is more accurately described by the three-band
model~\cite{perturbation_PRB2008}, here we consider a single-band TB
Hamiltonian, with the single active orbital being the Cu
$3d_{x^2-y^2}$ orbital at the Fermi level as the simplest case for the
proposed approach. In fact, recent studies showed that the one-band
Hubbard model suffices to describe the important features of the
low-energy physics in the cuprates~\cite{Medici09}.  We find that the
accuracy of the dopant-induced TB Hamiltonian parametrization suffers
from two sources of uncertainty, which are (i) the presence of
effective far Cu neighbor interactions in the homogeneous model
(undoped case) allowing for many alternative parameter sets and (ii)
the need to reduce the enormous number of adjustable parameters of the
dopant-induced TB model in order to perform the numerical
optimization. In view of these considerations we present two
alternative TB models and discuss their validity in terms of physical
arguments.

\section{Electronic structure calculations}

The electronic structure calculations for the parent compound {\BSCCO}
were performed with the reference crystal structure reported in
{Ref.~\onlinecite{Liang88}}.  {\BSCCO} crystallizes in the space group
$I4/mmm$, with a unit cell which we consider in the tetragonal
symmetry (see Ref.~\onlinecite{tetra?} for the discussion on the
structural supermodulation in {\BSCCO}), consisting of two identical
slabs of atoms, one shifted with respect to the other by a vector
$(a/2,a/2,c/2)$, where $a$ and $c$ are the lattice parameters of a
tetragonal unit cell.  Oxygen doping of this system was modeled by
introducing one extra interstitial oxygen atom into a surface
supercell consisting of eight primitive unit cells in the $xy$ plane.
We consider the surface supercell in order to reproduce the conditions
of a typical STM experiment on an O-doped {\dopedBSCCO} surface.  The
electronic structure calculations in the present work were performed
with the linearized-augmented-plane-wave (LAPW) basis, as implemented
in Wien2k \cite{Wien2k}.  The exchange and correlation effects were
treated within the generalized gradient approximation (GGA) as
implemented by Perdew, Burke and Ernzerhof~\cite{GGA}. Additional
details of the electronic structure calculations are given in Appendix
A.

\begin{figure*}
\includegraphics[width=\textwidth]{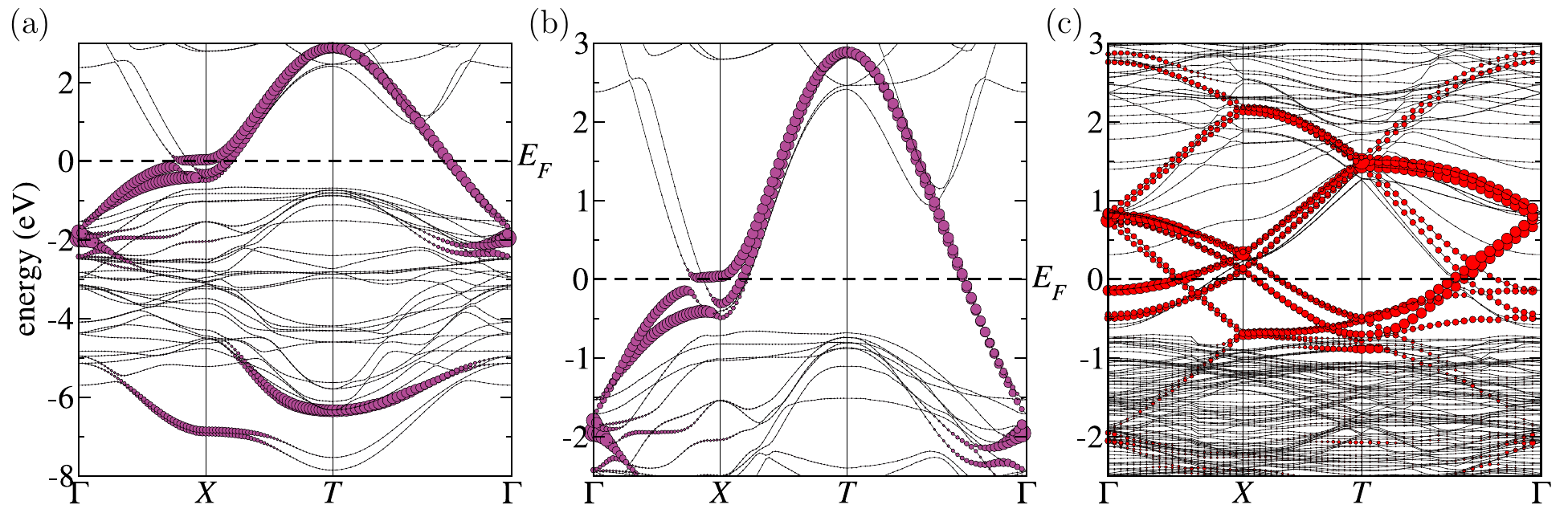}
\caption{(Color Online) (a) and (b) DFT electronic bandstructure of
  {\protect\BSCCO}  shown for different energy ranges.
   (c) Bandstructure of the O-doped
  {\protect\dopedBSCCO} supercell in the same energy window as (b). The weight of
  the Cu $3d_{x^2-y^2}$ character in the eigenvalues is proportional
  to the circles' size.}\label{bands}
\end{figure*}

In Fig.~\ref{bands}~(a) we present the electronic bandstructure of
{\BSCCO} in the energy window between $[-8\;\text{eV},3\;\text{eV}]$.
The two Cu $3d_{x^2-y^2}$ bands of the
parent compound are marked by circles with sizes proportional to the
$3d_{x^2-y^2}$ weight. They are rather dispersive, which is a typical
feature of the cuprate family~\cite{other_bands}.  The two Cu bands
crossing the Fermi level are the anti-bonding bands in the chemical
bonding between Cu atoms and the O atoms in the superconducting
layer. The bonding bands lie in the region between $[-8\;\text{eV},
 -5\;\text{eV}]$.
In the following  we concentrate on the energy window
$[-2.5\;\text{eV}, 3\;\text{eV}]$
near the Fermi level $E_F$ as indicated in
Fig.~\ref{bands}~(b).
{At the
$X=(\pi,0,0)$ point the Cu $3d_{x^2-y^2}$ bands show a strong overlap
with the Bi-O bands near the Fermi level, and at the $\Gamma=(0,0,0)$
point there is strong hybridization between the Cu $3d_{x^2-y^2}$ and
some of the lower-lying O $2p$ and Cu $3d$ bands.}

 Fig.~\ref{bands}~(c) presents the bandstructure of
the O-doped {\dopedBSCCO} supercell in the same energy range.  The bandstructure of
the supercell  is plotted along the same high
symmetry points in the Brillouin zone as those of the parent
compound. However, since the doped systems' unit cell is 8 times larger than
 the parent compound unit cell, its Brillouin zone shrinks and
the bands fold on top of each other so that effectively there are 16
Cu $3d_{x^2-y^2}$ bands crossing the Fermi level, where 16 is the
number of Cu atoms in the supercell slab.

Because the Brillouin zones of the parent compound and of the doped
supercell are defined differently, it is not straightforward to
compare their bandstructures.  In order to make such a comparison, we
recalculate the electronic structure of the parent compound in the
folded {Brillouin} zone. Once this is done, small but traceable
changes in the shape of the Cu $3d_{x^2-y^2}$ bands as a function of
oxygen doping become apparent as will be shown in the next
sections. We will dedicate the rest of the paper to the quantitative
evaluation of these changes in terms of hopping integrals of a
single-band tight-binding Hamiltonian as well as calculations of the
spin susceptibility and superconducting pair function.

\section{ Single-band tight-binding model}

\subsection{Parent compound}

A standard procedure to analyze the low-energy bandstructure features
of solid state systems is the tight-binding (TB)
 parametrization~\cite{Kandpal_09,Graser_susc}.
Since the symmetry of the parent compound {\BSCCO}
  unit cell is $I4/mmm$ and the
number of distinct hopping integrals (adjustable parameters) is small,
the TB parametrization in this case is straightforward
and has been already performed in previous studies~\cite{other_bands,Andersen_94}.
The complexity of the problem is dramatically increased when trying to
obtain the TB parameters for the doped supercell bands shown in
Fig.~\ref{bands}~(c): first, because the number of bands to be mapped
increases by a factor of 8, which means that, in order to optimize the
TB parameters,  a global minimum of a complex
mathematical function expressed by a $16\times16$ matrix needs to be found;
 second,
because the number of distinct hopping integrals is expected to rise
considerably as we increase the size of the unit cell and lower its
symmetry. Our strategy to overcome these complications is to use the
hopping integrals obtained for the parent compound Cu $3d_{x^2-y^2}$
bands as starting values for parametrizing the doped supercell bands.

An important point to consider is that the differences between the
Cu $3d_{x^2-y^2}$ bands of  the
parent and doped compounds  are
small, which is reasonable since we do not expect the interstitial
oxygen to have a drastic effect on the orbital overlap of its
neighboring Cu atoms. Therefore, the TB parameters for the
Cu $3d_{x^2-y^2}$ bands in the parent and doped compounds
must be obtained with a fitting error smaller than the dopant-induced
differences in the bandstructure, {\it i.e.}, the quality of both TB
mappings must be particularly high. For this reason we
consider the antibonding
Cu $3d_{x^2-y^2}$ bands over the entire energy range over which they disperse.
We find that, in order to {\it accurately} reproduce the DFT
Cu $3d_{x^2-y^2}$ bands of the parent compound [Fig.~\ref{bands}~(a)],
up to 13 Cu-Cu neighbors have to be included in the model TB Hamiltonian.
We should note, however, that among these various hopping integrals,
those representing hybridizations between the far neighbors are only
effective parameters; we are forced to include them in the single-band
model in order to describe those features of the Cu $3d_{x^2-y^2}$
bands that stem from the interaction of the Cu $3d_{x^2-y^2}$ orbital
with other Cu $3d$ orbitals and the O $2p$ orbitals.  For example,
near $\Gamma=(0,0,0)$ the Cu $3d_{x^2-y^2}$ bands are anomalously flat and
show an $\epsilon_k \propto$ $k^4$ behavior~\cite{Andersen_94},
where $k$ is the momentum in the {Brillouin zone}.  In order to reproduce this feature,
inclusion of higher harmonics are required in the model equations.

 The long range effective hoppings which arise from the mapping of
the complex band structure over a $\sim$ 5 eV range onto a single band model
are found to be much smaller than the short range hoppings, e. g., $t_{100}$ and $t_{110}$.
This strongly suggests that it will be difficult to identify a unique parameter set.
Alternative
single-band models can exist corresponding to different choices of
effective hopping paths.  
 We therefore discuss  two possible sets of
single-band TB Hamiltonian parameters (presented in Tables  
 \ref{parent_hoppings1} and \ref{parent_hoppings2} of Appendix B) to give a sense of how robust the
TB models can be.  Details of the construction of these models are found in the Appendix.
Both sets are indeed found to describe the Cu $3d_{x^2-y^2}$ bands in the parent compound
equally well.
In the following we will adopt the notation
\begin{itemize}
 \item TB1: 12 hopping parameters, 6 effective interlayer hoppings
 \begin{itemize}
   \item TB1$_{\rm undoped}$: Undoped parent compound.
   \item TB1$_{\rm loc.\;doped}$:  TB parameters are taken to vary in the vicinity of O dopant.
 \end{itemize}
 \item TB2: 13 hopping parameters, 4 effective interlayer hoppings
 \begin{itemize}
   \item TB2$_{\rm undoped}$: Undoped parent compound.
   \item TB2$_{\rm hom.\;doped}$:  Fit is obtained with homogeneous TB parameters over the entire supercell.
 \end{itemize}
\end{itemize}
A discussion of the tight-binding parameters describing the O-doped
system will be the subject of the next  Subsection.

\begin{figure}[h!]
\begin{center}
\includegraphics[width=0.35\textwidth]{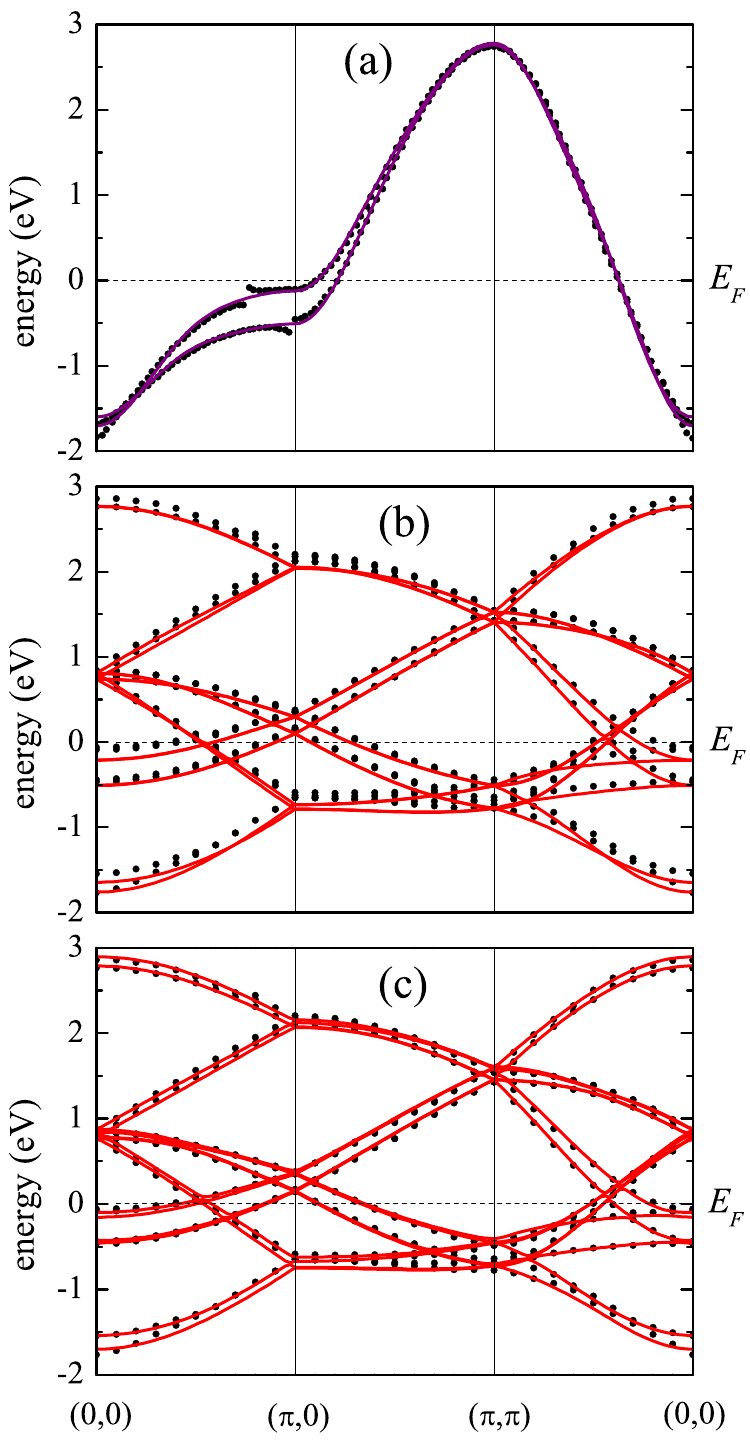}
\caption{(Color Online) Plot of the DFT calculated Cu $3d_{x^2-y^2}$ bands
(dots) and the
  TB1 Hamiltonian spectrum (lines) for {\BSCCO} and {\dopedBSCCO}:
 (a) comparison of the {\BSCCO} electronic structure to the TB1$_{\rm undoped}$
model;
  (b) comparison of the {\dopedBSCCO} electronic structure to the
  TB1$_{\rm undoped}$ model  plotted in the folded Brillouin
  zone; (c) comparison of the {\dopedBSCCO} electronic structure to
  the TB1$_{\rm loc.\;doped}$ model
   (see Fig.~\ref{local_hop}). High symmetry points are given by
   $(k_x,k_y)$ only, $k_z=0$; thus $(0,0)=\Gamma$, $(\pi,0)=X$, $(\pi,\pi)=T$. }\label{bands_f1}
\end{center}
\end{figure}

In 
Fig.~\ref{bands_f1}~(a) the TB1$_{\rm undoped}$ Hamiltonian spectrum in the
full Brillouin zone is compared to the DFT bands of the parent compound.
In Fig.~\ref{bands_f1}~(b), the spectrum of the TB1$_{\rm undoped}$ Hamiltonian  is
replotted in the folded Brillouin zone and compared with the Cu
$3d_{x^2-y^2}$ bands of the doped supercell. The slight differences
between the shape of the doped supercell DFT bands and of the TB
 bands of the parent compound are due to the presence of the
interstitial oxygen, which displaces the neighboring Cu atoms and thus
modifies their corresponding overlap integrals.  It is possible to
quantify these effects by fine-tuning the parameters of the TB
Hamiltonian such that the
supercell bands are reproduced, as we will show below.
\begin{figure}[h!]
\begin{center}
\includegraphics[width=0.35\textwidth]{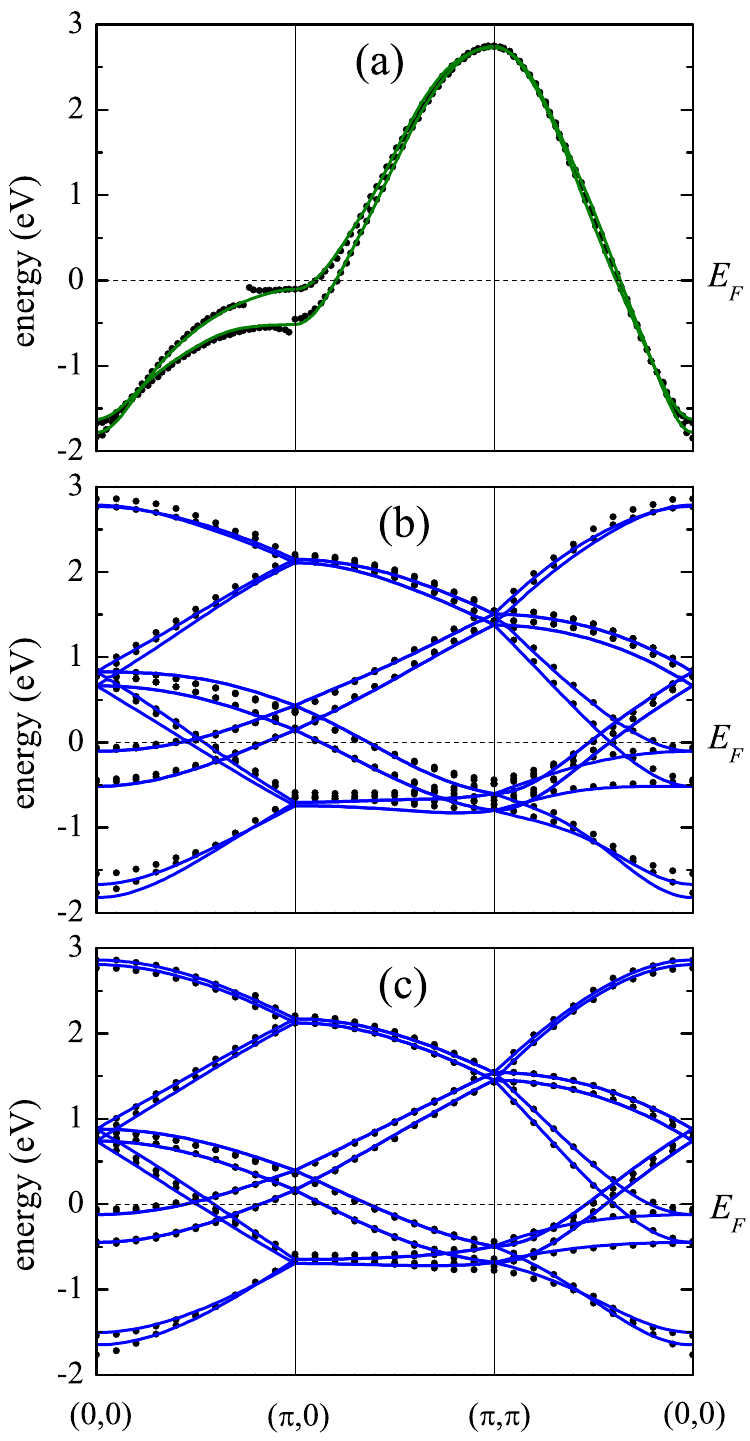}
\caption{(Color Online) Plot of DFT calculated Cu $3d_{x^2-y^2}$ bands (dots) and the
  TB2 Hamiltonian spectrum (line) for {\BSCCO} and {\dopedBSCCO}:
(a) comparison of the {\BSCCO} electronic structure to the TB2$_{\rm undoped}$
model;
  (b) comparison of the {\dopedBSCCO} electronic structure to the
  TB2$_{\rm undoped}$ model  plotted in the folded Brillouin
  zone; (c) comparison of the {\dopedBSCCO} electronic structure to
  the homogeneous TB2$_{\rm hom.\;doped}$ model.
   (
  Table~\ref{parent_hoppings2} second row).}\label{bands_f2}
\end{center}
\end{figure}

 In Fig.~\ref{bands_f2}~(a), we now show the 
corresponding
TB2$_{\rm undoped}$ Hamiltonian
spectrum  in the full Brillouin
zone in comparison to the DFT bands of the parent compound. In
Fig.~\ref{bands_f2}~(b), the same TB bands are plotted in the folded Brillouin zone, together with the
DFT bands of the doped supercell.
Comparing  the TB1$_{\rm undoped}$  and
TB2$_{\rm undoped}$ model parameters for the parent compound
(Tables~\ref{parent_hoppings1} and~\ref{parent_hoppings2} of Appendix B), we
observe that the TB2$_{\rm undoped}$ model has several
features that can be considered an advantage in terms of the physics
that the model implies; this   model includes only four
hopping integrals between the CuO$_2$ layers, $t_{00z}$, $t_{11z}$,
$t_{21z}$ and $t_{33z}$, whose relevance can be justified either by
the close proximity of the two Cu atoms ($t_{00z}$) or by the
presence of a Ca atom along the Cu-Cu connection mediating electron
hopping ($t_{11z}$, $t_{21z}$ and $t_{33z}$). In the TB1$_{\rm undoped}$
model, on the other hand, the mechanism of some of its six inter-layer
interactions is not as clear. Furthermore,  one would rather
expect the contribution of interacting far Cu neighbors within a CuO$_2$
layer to be more important as considered
 in the TB2$_{\rm undoped}$ model.

In the next section, we will discuss the results of the derivation of
the TB Hamiltonian for the O-doped {\dopedBSCCO} supercell, obtained
when  either the parameter set in TB1$_{\rm undoped}$
(Table~\ref{parent_hoppings1}) or  TB2$_{\rm undoped}$
(Table~\ref{parent_hoppings2}) are used as initial values for mapping
the supercell DFT bands.

\subsection{  O-doped supercell}

In order to describe the DFT Cu $3d_{x^2-y^2}$ bands of the doped
supercell [Fig.~\ref{bands}~(c)] within  a TB model, one has to
construct a Hamiltonian similar to Eq.~(\ref{Ham}), which will be now
represented by a 16$\times$16 matrix, according to the number of Cu
atoms in the supercell (8 atoms in each CuO$_2$ layer). Since the
presence of the interstitial oxygen introduces inhomogeneities in the
system, the number of distinct model parameters for the supercell is
not defined by simply the number of parameters of the corresponding
parent unit cell (which is 12 and 13 for the models of
Table~\ref{parent_hoppings1} and Table~\ref{parent_hoppings2},
respectively, plus the on-site energy $\mu$), but increases considerably.
 For instance, even by taking into account the mirror plane symmetry,
 there are still 238 parameters in the supercell TB
Hamiltonian based on the TB1$_{\rm undoped}$ {\it ansatz}.
  Technically, it is impossible to find a unique and
unambiguous set of parameters by performing an optimization of such a huge number of
parameters, especially since  our aim is to capture the slight
differences between the bands of the parent compound and the doped
compound [see Fig.~\ref{bands_f1}~(b) and Fig.~\ref{bands_f2}~(b)].
{One way to proceed
would be to approximate the hopping integrals that become distinct in
the supercell due to the inhomogeneity introduced by the dopant by
their average values.}
 In this ``averaged'' homogeneous TB model for
the supercell, there would be as many parameters as in the
corresponding model for the parent compound, and their optimization
would be simple; an example is given in Table~\ref{parent_hoppings2}
(TB2$_{\rm hom.\;doped}$ row). With such an approach, however, the most interesting
physics concerning {\it local} effects due to the dopant is left out.

The exact knowledge of how the Cu on-site energies and the most
relevant Cu-Cu hopping integrals $t_{100}$ and $t_{110}$ are modified
near the dopant is very important for understanding the dopant-induced
effects on the local spin superexchange coupling \cite{Anderson_87,Super_theory1,
  Super_theory2}, which is related to the size of the local
superconducting gap in cuprates~\cite{perturbation_PRB2008}.
Therefore, in order to be able to study the local variations in the
model parameters, we propose the following approximate treatment of
the problem. We assume that the on-site energies and hopping integrals
most affected by the dopant are those that are nearest to the dopant,
and we concentrate on the  largest TB model parameters, such as
$\mu$, $t_{100}$ and $t_{110}$. Then, the supercell TB Hamiltonian is
optimized by adjusting the selected parameters, while for the rest,
{\it i.e.}, the effective far Cu neighbor interactions, their initial
values are preserved.
An illustration of the procedure in the case of the
TB1$_{\rm undoped}$ model is shown in Fig.~\ref{local_hop}
and is worked out in detail in Appendix C.
It turns out that this approach gives sensible
results when the TB supercell model is built upon the
 TB1$_{\rm undoped}$ model (Table~\ref{parent_hoppings1})
and gives counterintuitive results
when the TB2$_{\rm undoped}$ model for the parent compound
(Table~\ref{parent_hoppings2}) is used.
In Fig.~\ref{bands_f1} (c) we show
the good agreement between the energy spectrum of this
TB1$_{\rm loc.\;doped}$ Hamiltonian
with the DFT bands of \dopedBSCCO.

\begin{figure}[tb]
\includegraphics[width=0.45\textwidth]{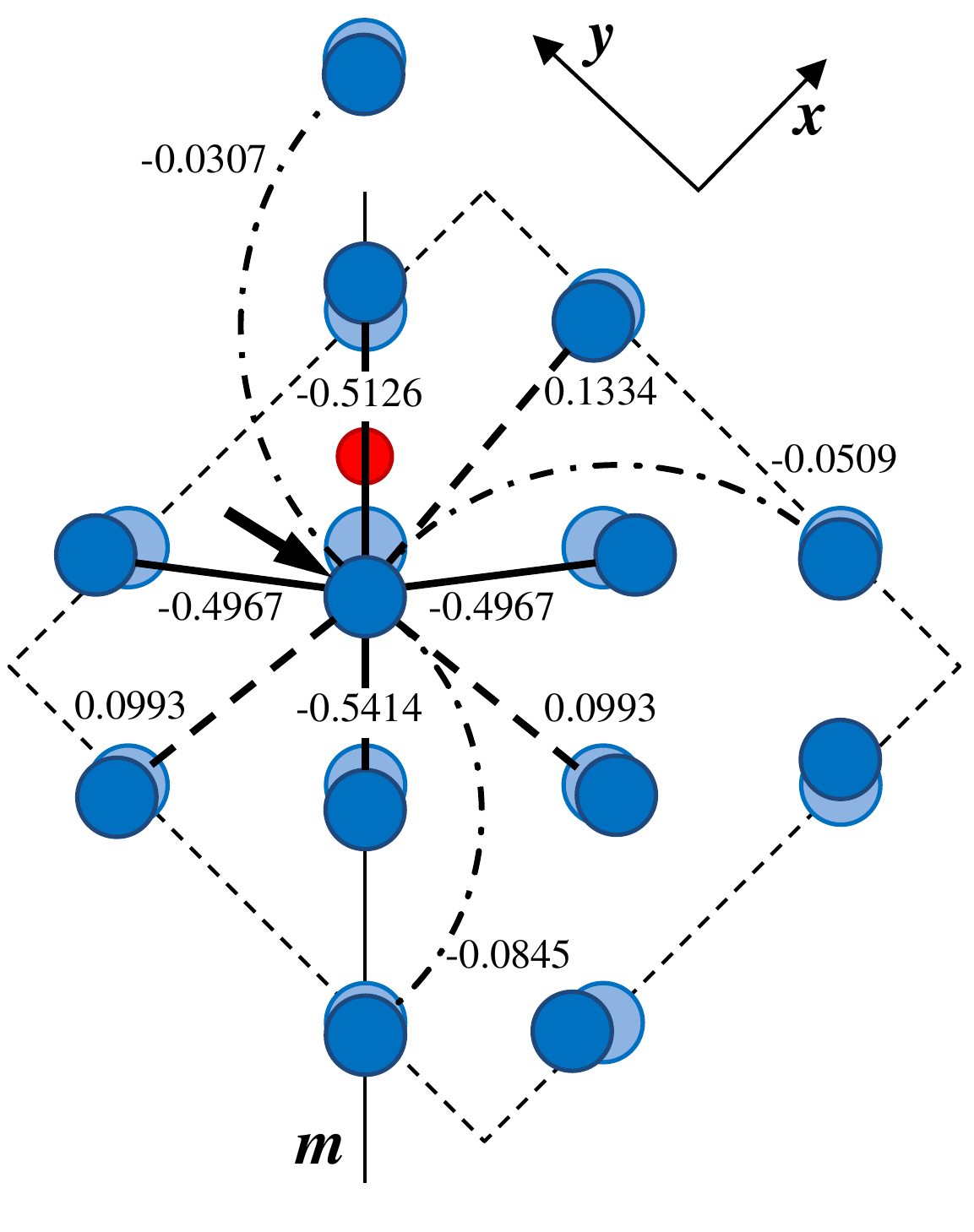}
\caption{(Color Online) The schematic lattice of dopant-displaced Cu
  atoms in the oxygen-doped supercell of {\dopedBSCCO}.  The smallest
  circle represents the interstitial oxygen atom, and the larger
  circles stand for Cu atoms. Darker color is used for Cu atoms in the
  CuO$_2$ layer closest to the interstitial oxygen.  The Cu-Cu bonds
  that correspond to the hopping integrals of the $t_{100}$ type are
  represented by solid lines, the hopping integrals of the $t_{110}$
  type by dashed lines, and the hopping integrals of the $t_{200}$
  type by dash-dotted lines. The numbers over the bonds stand for the
  optimized values of corresponding hopping integrals of the doped
  supercell TB model based upon the nearest-neighbors parent compound
  model (see text). The optimized value of the on-site energy of this
  model for the Cu in the next-nearest CuO$_2$ layer (the light Cu
  atom symbols) is $\mu=0.4445$ eV. The six optimized on-site energy
  values for Cu in the nearest CuO$_2$ layer (the dark Cu atom
  symbols) are $\mu=0.5757$ eV, $0.5057$ eV, $0.5341$ eV, $0.5151$ eV,
  $0.4930$ eV, and $0.5186$ eV. $\mu= 0.5757$ eV corresponds to the Cu
  atom which is displaced most by the dopant and is marked with an
  arrow.  $m$ labels the mirror plane.}\label{local_hop}
\end{figure}

To conclude this section, we present the {\it homogeneous} TB model
for the doped supercell TB2$_{\rm hom.\;doped}$ based on the TB2$_{\rm undoped}$
model (see  Table~\ref{parent_hoppings2} and
Fig.~\ref{bands_f2}~(c)). Even though the homogeneous model
does not reflect the local dopant-induced effects, it can still be
useful as it provides the information on how the model parameters
change on average.  For instance, it is interesting to observe that
the average ratio $t_{110}/t_{100}$ increases in the doped supercell
compared to that in the parent compound (from 0.2097 to 0.2249); this
might suggest a possible increase of the superconducting transition
temperature upon doping, in analogy with the observation within
the cuprate family, that the materials characterized by a larger
$t_{110}/t_{100}$ ratio have higher transition
temperatures~\cite{ratioVSTc}. Unlike the situation with the
mapping approach previously discussed, which aimed at capturing local
physics, the parameters of the homogeneous Hamiltonian demonstrate the
same   behavior  upon doping regardless of the parent
compound TB model (TB1$_{\rm undoped}$ and TB2$_{\rm undoped}$)
chosen as a starting point for mapping the doped supercell
electronic structure.

\section{Spin fluctuation pairing}
\subsection{ Spin susceptibility}

In the following we calculate the magnetic spin susceptibilities for
the TB models obtained previously, namely, (i) the TB1$_{\rm undoped}$ (Table~\ref{parent_hoppings1}), (ii) the
TB2$_{\rm undoped}$ (Table~\ref{parent_hoppings2} first row),
(iii) the inhomogeneous doped supercell model TB1$_{\rm loc.\;doped}$ (Fig.~\ref{local_hop})
and (iv) the homogeneous doped supercell model TB2$_{\rm hom.\;doped}$
(Table~\ref{parent_hoppings2} second row).  The
spin susceptibility is derived within
the Matsubara Green's functions formalism~\cite{Mahan} from the
non-interacting Green's functions.
In a general formulation, the spin
susceptibility is a function of four orbital indices, $(\chi_{\rm s})^{pq}_{st}$,
which, in the considered case of a single orbital but multiple atoms in a unit cell,
refer to the orbitals on different atoms.
For the non-interacting case, the spin susceptibility $(\chi_{\rm s})^{pq}_{st}$
is equivalent to the charge susceptibility $(\chi_{\rm c})^{pq}_{st}$,
$(\chi_{\rm s})^{pq}_{st}=(\chi_{\rm c})^{pq}_{st}\equiv\chi^{pq}_{st}$,
and is given by~\cite{Graser_susc}:
\begin{equation}\begin{split}
\chi^{pq}_{st}({\bf q},\omega)=-&\frac{1}{NN_{{\bf k}}}
\sum_{{\bf k},\mu\nu} \left[f(E_{\nu}({\bf k}+{\bf q}))-f(E_{\mu}({\bf k}))\right] \\&\times
\frac{a^s_{\mu}({\bf k})a^{p*}_{\mu}({\bf k})a^q_{\nu}({\bf k}+{\bf q})
a^{t*}_{\nu}({\bf k}+{\bf q})}{\omega+E_{\nu}({\bf k}+{\bf q})-E_{\mu}
({\bf k})+{\rm i}0^+}
\,.
\end{split}
\end{equation}

In this expression, indices $s$, $p$, $q$ and $t$ refer to the $N$ Cu atoms in
the unit cell and run from 1 to $N$ while indices $\mu$ and $\nu$
distinguish the $N$ eigenvalues $E({\bf k})$ of the diagonalized TB
Hamiltonian.  The matrix elements $a^s_{\mu}({\bf k})$ are the
components of the eigenvectors of the TB Hamiltonian.  The integration
over the Brillouin zone has been replaced by a sum over a sufficiently
large number $N_{{\bf k}}$ of $k$-points. $f(E)$ is the Fermi-Dirac
distribution function. In the following, we will focus on the static
non-interacting spin susceptibility $\chi_{\rm S}({\bf q})$,
\begin{equation}
\chi_{\rm S}({\bf q}) = \frac{1}{2}
\sum_{sp}\chi_{ss}^{pp}({\bf q},\omega=0),
\end{equation}
and examine its behavior in the four cases of interest
along the main symmetry directions in the Brillouin zone.

\begin{figure}[tb]
\begin{center}
 \includegraphics[trim = 5mm 0mm 6mm 6mm,width=0.45\textwidth]{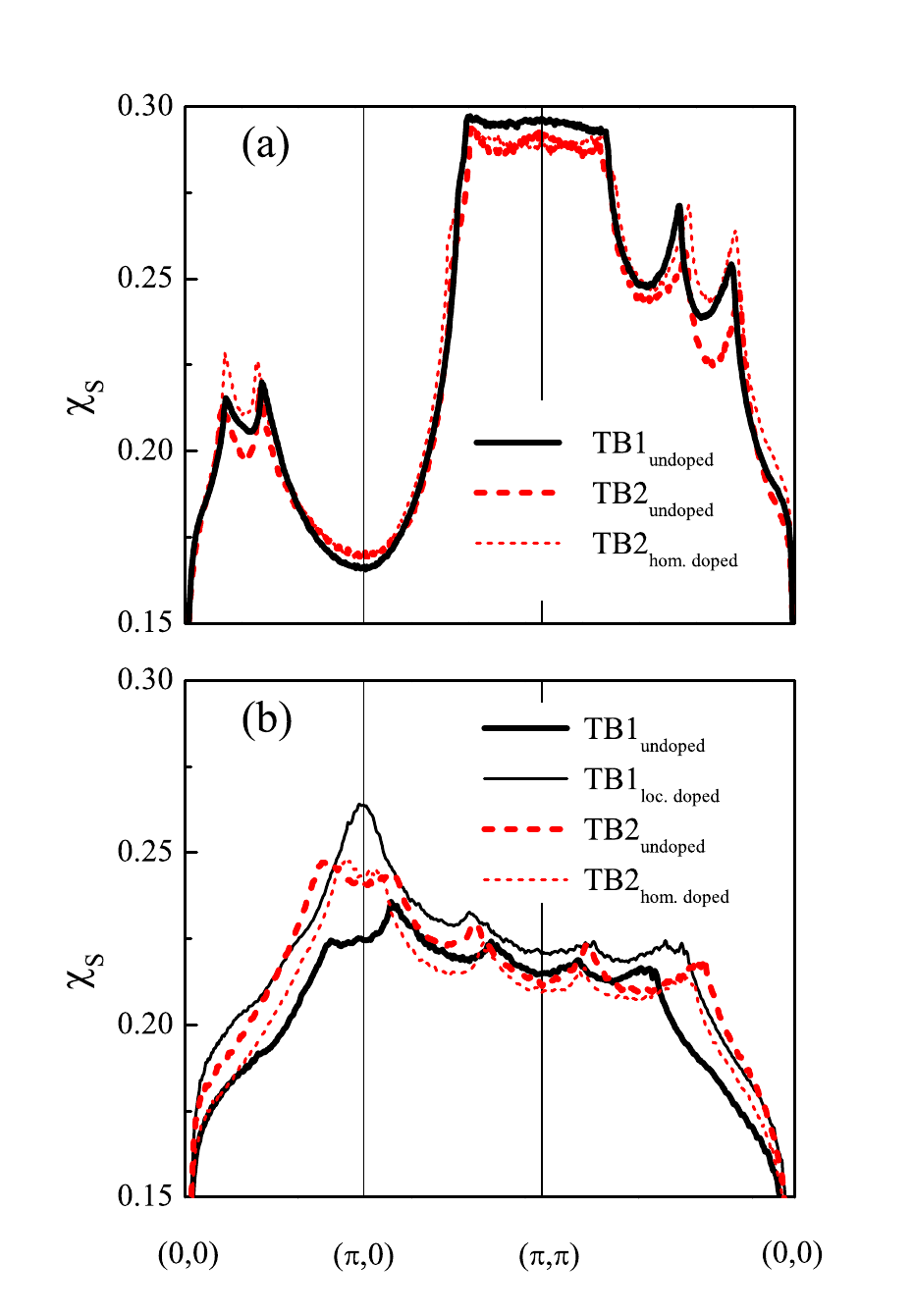}
\caption{The static spin susceptibility of (a) the two parent compound
  TB models TB1$_{\rm undoped}$ and TB2$_{\rm undoped}$
 and the homogeneous doped supercell model
  TB2$_{\rm hom.\;doped}$, plotted in the full Brillouin zone, and (b) the two parent
  compound TB models TB1$_{\rm undoped}$ and TB2$_{\rm undoped}$ and the inhomogeneous {TB1$_{\rm loc.\;doped}$} and
  homogeneous TB2$_{\rm hom.\;doped}$ doped supercell models, plotted in the folded
  Brillouin zone.}\label{susc}
\end{center}
\end{figure}

The static spin susceptibilities $\chi_{\rm S}({\bf q})$ of the parent
compound calculated with the TB1$_{\rm undoped}$  model of
Table~\ref{parent_hoppings1} (bold black line) and with the
TB2$_{\rm undoped}$ model of Table~\ref{parent_hoppings2} (bold dashed line)
are plotted in Fig.~\ref{susc}~(a).  The two susceptibilities show
similar features with double peaks along
$(0,0,0)-(\pi,0,0)$ and $(\pi,\pi,0)-(0,0,0)$ directions and a broad
plateau at $(\pi,\pi,0)$.  These similarities can be understood
in terms of the fact that
 most important parameters in the two TB models ($t_{100}$, $t_{110}$, etc.)
have close values. In this respect,
it is not surprising that the spin susceptibility calculated with the
averaged TB parameters of the homogeneous Hamiltonian TB2$_{\rm hom.\;doped}$
for
{\dopedBSCCO} (thin dashed line in Fig.~\ref{susc}~(a)) qualitatively
reproduces the same behavior as  the TB models for the
parent compound.

We next calculate the spin susceptibility with
the {\it inhomogeneous} TB Hamiltonian
for the doped {\dopedBSCCO} supercell  TB1$_{\rm loc.\;doped}$.
Of course in this case the spin susceptibility  must be
calculated with the full supercell
$(16\times 16)$ Hamiltonian matrix  and accordingly is
defined in the
folded Brillouin zone.  Fig.~\ref{susc}~(b) shows the spin
susceptibility calculated with the inhomogeneous TB1$_{\rm loc.\;doped}$ model (thin
black line) and, for comparison, the two parent compound
susceptibilities replotted in the folded Brillouin zone (as in
Fig.~\ref{susc}~(a), bold black and bold dashed lines). Within the
inhomogeneous model, a pronounced peak in the spin susceptibility
evolves at $(\pi,0,0)$ upon doping whereas in both parent compound
susceptibilities this region is featured by a shallow minimum in
between two asymmetrical peaks located at some distance from
$(\pi,0,0)$.  A peak in the spin susceptibility of a non-interacting
system can transform into a divergence indicating magnetic
instabilities and possible ordering, when the interparticle
interactions are switched on.  $(\pi,0,0)$ corresponds to a
commensurate antiferromagnetic striped order with period $2\sqrt{2}a$,
with stripes along the (110) direction of the parent compound unit
cell.

\subsection{ Superconducting gap function}
We consider now the models TB1$_{\rm undoped}$ and TB1$_{\rm
  loc.\;doped}$ in order to analyze the superconducting properties of
the undoped and doped {\dopedBSCCO}, respectively.  We calculate the
pairing vertex by assuming that superconductivity in the high-$T_c$
cuprates is driven by the exchange of spin and charge
fluctuations~\cite{Bickers_89}. The many-body effects of the Coulomb
interaction are here treated within the random phase approximation
(RPA).

In order to calculate the pairing vertex, the RPA charge and spin
susceptibilities, $\chi_{\rm c}^{\rm RPA}({\bf q},\omega)$ and $\chi_{\rm s}^{\rm RPA}({\bf q},\omega)$, are required.
They can be obtained from the non-interacting susceptibility $\chi({\bf q},\omega)$
in the form of Dyson-type equations as
\begin{equation}
 (\chi_{\rm c}^{\rm RPA})^{pq}_{st} =
\chi^{pq}_{st}-\sum_{uvwz}(\chi_{\rm c}^{\rm RPA})^{pq}_{uv}
(U^{\rm c})^{uv}_{wz}\chi^{wz}_{st}
\end{equation}
and
\begin{equation}
 (\chi_{\rm s}^{\rm RPA})^{pq}_{st} =
\chi^{pq}_{st}+\sum_{uvwz}(\chi_{\rm s}^{\rm RPA})^{pq}_{uv}
(U^{\rm c})^{uv}_{wz}\chi^{wz}_{st}.
\end{equation}
For a single-band model, only the diagonal $U^{\rm c}$
and $U^{\rm s}$ matrices' components are non-zero:
\begin{equation}
 \left(U^{\rm c}\right)^{ii}_{ii} = U,\quad\left(U^{\rm s}\right)^{ii}_{ii} = U,
\end{equation}
where $U$ is the strength of the on-site intra-band Coulomb repulsion between electrons.
The singlet pairing vertex is then given by
\begin{eqnarray}
 \Gamma^{pq}_{st}({\bf k},{\bf k}',\omega)
&=&\left[
\frac{3}{2}U^{\rm s}
\chi^{\rm RPA}_{\rm s}({\bf k}-{\bf k}',\omega)U^{\rm s}
+\frac{1}{2}U^{\rm s}\right.\\
&&-\left.\frac{1}{2}U^{\rm c}
\chi^{\rm RPA}_{\rm c}({\bf k}-{\bf k}',\omega)U^{\rm c}
+\frac{1}{2}U^{\rm c}\right]^{tq}_{ps}.\nonumber
\end{eqnarray}
The scattering of a Cooper pair from the state $({\bf k},-{\bf k})$ to the state $({\bf k}',-{\bf k}')$
on the Fermi surface is determined by the projected interaction vertex
\begin{eqnarray}
 \Gamma({\bf k},{\bf k}',\omega) &=& \sum_{stpq}
a^{t}_{\nu}(-{\bf k})a^{s}_{\nu}({\bf k})\Gamma^{pq}_{st}({\bf k},{\bf k}',\omega)
\nonumber\\
&\times&a^{p,*}_{\nu'}({\bf k}')a^{q,*}_{\nu'}(-{\bf k}'),
\end{eqnarray}
where indices $\nu$ and $\nu'$ refer to the eigenvectors of the
TB Hamiltonian
with the corresponding energy eigenvalues close to the Fermi level.
As the strength of the pairing interaction is defined
by a frequency integral of the imaginary part of $\Gamma({\bf k},{\bf k}',\omega)$
weighted by $\omega^{-1}$, it is sufficient to consider the real
part of $\Gamma({\bf k},{\bf k}',\omega=0)$ according to the Kramers-Kronig relation:
\begin{equation}
 \int_0^{\infty}d\omega\frac{{\rm Im}[\Gamma({\bf k},{\bf k}',\omega)]}
{\pi\omega} = {\rm Re}[\Gamma({\bf k},{\bf k}',\omega=0)].
\end{equation}

If the superconducting gap is decomposed into an amplitude
$\Delta$ and a normalized gap function
$g({\bf k})$, the latter can be evaluated from the
following eigenvalue equation
\begin{equation}
 -\oint \frac{d {\bf k}'_{||}}{2\pi}
\frac{1}{2\pi v_F({\bf k}')}\Gamma^{\rm symm}({\bf k},{\bf k}')g({\bf k}')=\lambda g({\bf k}).
\label{eigenvalue}
\end{equation}
Here,
\begin{equation}
 \Gamma^{\rm symm}({\bf k},{\bf k}') = \frac{1}{2}{\rm Re}
\left[\Gamma({\bf k},{\bf k}',0)+\Gamma({\bf k},-{\bf k}',0)\right]
\end{equation}
is the symmetric part of the full interaction and
\begin{equation}
 v_F({\bf k})=|\nabla_{\bf k} E_{\nu}({\bf k})|
\end{equation}
is the Fermi velocity at point ${\bf k}$ on the Fermi surface.
The largest eigenvalue $\lambda$ of Eq. (\ref{eigenvalue}) determines the
superconducting transition temperature and its corresponding
eigenfunction $g({\bf k})$ has the symmetry of the gap.

  We have solved the eigenvalue problem (\ref{eigenvalue}) for
the undoped and doped {\dopedBSCCO} models in the folded
Brillouin zone of the supercell. The folded Brillouin zone has been
considered in both cases in order to ensure that the eigenvalue equations
are constructed
under the same conditions, which is important when the resulting pairing
strengths are compared. The calculations have been performed for the temperature
$T=0.01$ eV and we considered  Coulomb repulsion $U$ values that range from 1.00 eV
to 1.66 eV. Note that these values represent renormalized values of the Hubbard $U$
appropriate for RPA treatments and are smaller than bare $U$'s~\cite{BulutScalapino}.
\begin{figure}[tb]
\begin{center}
\subfigure {\includegraphics[trim = 5mm 0mm 6mm 0mm, clip,width=0.45\textwidth]{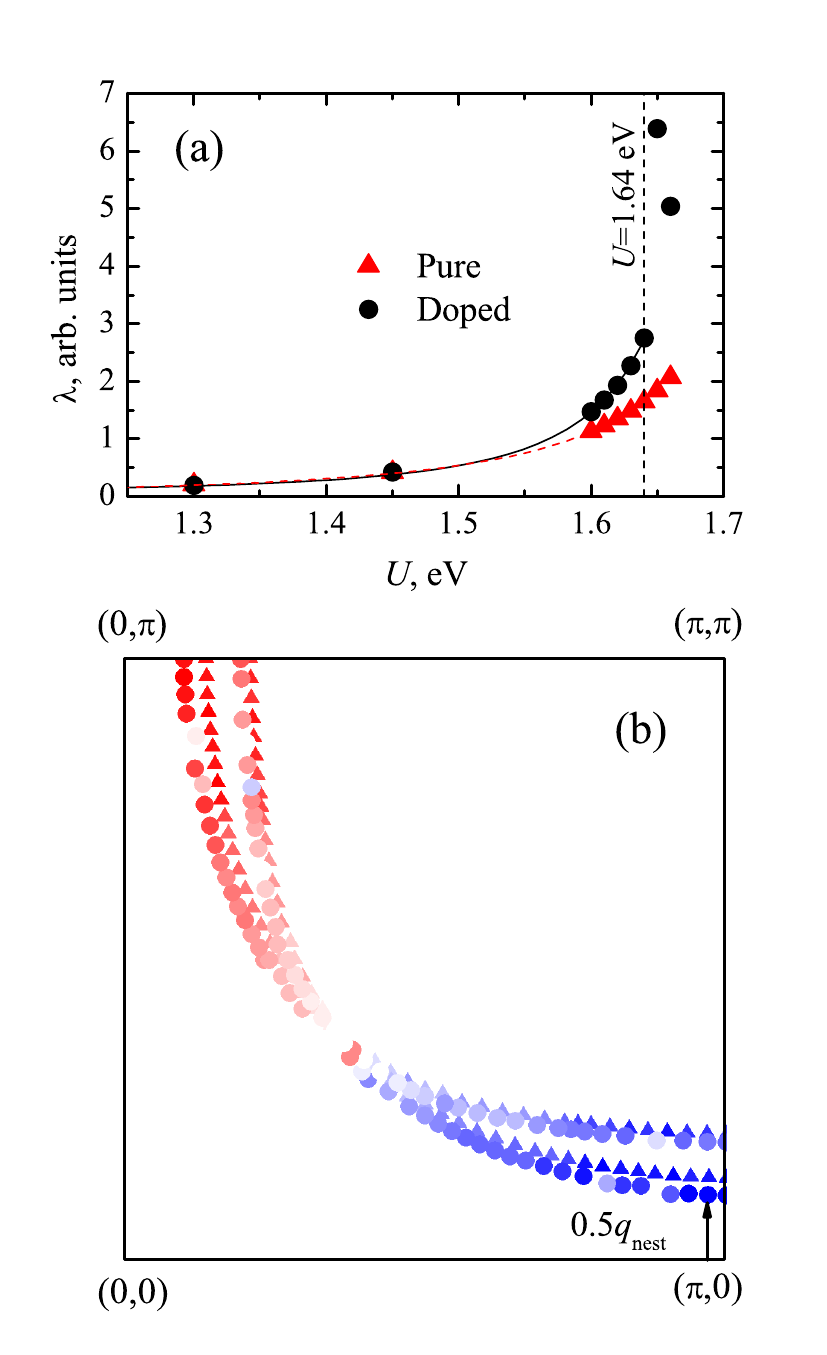}}
\caption{(Color Online) (a) The pairing strength $\lambda$ for the undoped (triangles) and
doped (circles) {\dopedBSCCO} TB models as a function of Coulomb repulsion $U$.
(b) The superconducting gap function $g({\bf k})$
on the $k$-point mesh at the Fermi surface of the undoped {\BSCCO} unit cell
for the undoped (triangles) and doped (circles) models. The red (blue)
color represents positive (negative) $g({\bf k})$ values, and the
intensity of the color is proportional to the absolute value of $g({\bf k})$.
Half the nesting vector $q_{\rm nest}$ is shown by an arrow.}\label{gap}
\end{center}
\end{figure}

 We find that the doped {\dopedBSCCO} model is characterized by a
larger value of the pairing strength $\lambda$ compared to that of the undoped model.
The pairing strengths for the two models are presented in Fig.~\ref{gap}~(a)
as a function of $U$. Below $U\sim1.5$ eV the two $\lambda$
values are almost equal, but
at larger $U$ values the pairing strength for the doped model grows
faster and diverges at $U=1.65$ eV.

  Fig.~\ref{gap}~(b)
displays the gap functions $g({\bf k})$ of the undoped and
doped models, corresponding to the leading eigenproblem solutions $\lambda$ of Fig.~\ref{gap}~(a)
at $U=1.64$ eV.
The result from Eq. (\ref{eigenvalue})  $g({\bf k})$ is defined on the mesh
of $k$-points at the Fermi surface of the folded Brillouin zone.
In Fig.~\ref{gap}~(b),
the $k$-point mesh was unfolded to the Brillouin zone of the undoped
compound unit cell in order to allow a comparison of the Fermi surface behavior
for the two systems with experiment.
 One should note that in the case of the doped supercell such an unfolding
is, strictly speaking, not allowed and results in a
tearing of the Fermi surface. Yet, since
the symmetry lowering effects caused by a dopant are small,
 the unfolding
in this case is a reasonable approximation.
In particular, the unfolded way of presenting $g({\bf k})$
allows us to observe that the symmetry of the undoped model $g({\bf k})$ is $d_{x^2-y^2}$ and
that upon doping it is roughly preserved, though slightly distorted. We also
note the characteristic reduction of the norm of the nesting wave vector $q_{\rm nest}$,
associated with the occurrence of the spin density wave state, in the doped
case. This behavior is in agreement with ARPES results for {\dopedBSCCO}
~\cite{Wise_08}.

  The gap equation calculations presented here show that the TB model
derived for the O-doped {\dopedBSCCO}, TB1$_{\rm loc.\;doped}$
shows an enhanced superconducting pairing compared to the parent compound model TB1$_{\rm undoped}$.
This model also demonstrates the appearance of the $(\pi,0,0)$ peak in the non-interacting
static spin susceptibility. These two features of the doped model prove the
suggested important role of local crystal and electronic structure inhomogeneities due to
doping for the local superconducting properties of {\dopedBSCCO}.

\section{Conclusions}
A single-band TB model parametrizing the Cu $3d_{x^2-y^2}$ bands
 for the high-temperature superconductor
{\dopedBSCCO} was derived from DFT electronic structure calculations.
In particular we analyzed the changes in the TB model parameters induced
by the dopant oxygen atom.
We found that an accurate quantitative analysis of the dopant-induced changes
of the electronic structure of {\dopedBSCCO} requires
high-quality TB parametrizations. The required quality of the modelling was achieved by
including effective far-neighbor interactions that increased the
number of TB parameters in the parent compound model up to 14. Two
possible models for the parent compound were proposed and compared,
one based on the nearest-neighbors interactions and the second, based on
 presumably physically justified hybridization paths with an
emphasis on intra-layer ones. These two parent compound TB models were
used for constructing the TB models for the doped supercell. The
nontrivial problem of mapping the doped supercell bandstructure was
treated by approximating the doped supercell TB Hamiltonian either by
a homogeneous one, with averaged parameters, or by one where only
certain selected parameters were adjusted to map the DFT bands. The
more promising latter approach gave results consistent with physical
intuition when applied to the nearest-neighbors parent compound TB
model.

 The static spin susceptibility
calculated with  the doped supercell TB1$_{\rm loc.\;doped}$ model  possesses qualitatively new
features, namely, a pronounced peak at the $X=(\pi,0,0)$ point in the folded
Brillouin zone, compared to the susceptibility of the parent compound.
 This change in the susceptibility was shown to lead to a
significant enhancement of $d$-wave pairing in the same {\dopedBSCCO} model.
Depending on the value of the interaction parameter $U$ chosen, a modulation
of the coupling constant $\lambda$ of order $30\%$ required to explain the STS
phenomenology\cite{Nunner} is easy to obtain.
While this is not a strictly local calculation, it is a strong indication that
-- within weak coupling theory --
the local electronic structure change caused by the interstitial O dopant in 
{\dopedBSCCO} in fact enhances the pairing locally.  A full local calculation of 
the inhomogeneous spin fluctuation pairing interaction is clearly desirable.

Other extensions of the current calculation would be useful.
Since it is known that the three-band model describes the magnetic
properties of cuprates better, it would be worthwhile to try to extend
the approach presented in this paper by performing a parametrization of
the three-band TB Hamiltonian and studying the effects of doping on
this model's parameters. In this case, one expects, on the one hand, a
growth of the number of model parameters due to new degrees of
freedom. On the other hand, the number of effective far-neighbor
interactions should decrease as many effects due to hybridization
between Cu and O orbitals are now incorporated in the model itself.
One could also consider other {\it multi-band} TB models by taking
more Cu and O orbitals, besides the Cu $3d_{x^2-y^2}$, O $2p_x$ and O
$2p_y$ orbitals, into account.  The multi-band models can be very
efficiently parametrized within the Slater-Koster
formalism~\cite{Slater-Koster}; moreover, the formalism allows direct
calculation of overlap integrals in the doped compound as a function
of relative atomic positions instead of parametrization by fitting.
This will be a subject of further investigations.

\section{Acknowledgments}
We would like to thank   J.C. Davis,  T. P. Devereaux,
 K. McElroy,  T. Saha-Dasgupta,
and Y.-Z. Zhang for useful discussions and gratefully acknowledge
financial support from the Deutsche Forschungsgemeinschaft
through the SFB/TRR 49 program and through TRR~80~(SG) and from the Helmholtz
Association through HA216/EMMI. PJH was supported by DOE DE-FG02-05ER46236
and HPC was supported by DOE/BES DE-FG02-02ER45995.

\setcounter{secnumdepth}{1}
\appendix

\section{Details of the electronic structure calculations}

 The {\dopedBSCCO} supercell is constructed in the
following way. It is
rotated with respect to the primitive unit cell by $45^{\circ}$ around the
$z$-axis and extended along the new $x$- and $y$-axes such that its $xy$ dimensions
are $(2\sqrt{2}a\times2\sqrt{2}a)R45^{\circ}$, where $a$ is the $x$ dimension of the pure {\BSCCO} unit cell. Then only one slab of the
two {\dopedBSCCO} slabs is considered.
 In the $z$ direction, the slabs are separated by
approximately 15~\AA\; of vacuum in order to exclude any interaction
between the marginal Bi atoms of adjacent slabs.  The exact atomic
positions inside the supercell slab were determined by He {\it et al.}
\cite{He_2006}, who performed structural optimization calculations
with the Vienna {\it ab initio} simulation program ({\sc VASP}) within
the local density approximation.

 Calculations for the parent
compound were carried out with an energy cut-off for the basis set
size given by $R_{\rm MT}K_{\rm max}=5.50$ ($R_{\rm MT}$ is the
smallest muffin tin radius and $K_{\rm max}$ is the maximal lattice
vector considered).
{The muffin tin radii for the different atoms in the unit cell were
the following: $R_{\rm MT}$(Bi) = 1.88~bohr, $R_{\rm MT}$(Sr) = 2.22~bohr, $R_{\rm MT}$(Ca) = 2.17~bohr,
$R_{\rm MT}$(Cu) = 1.82~bohr, and $R_{\rm MT}$(O) = 1.61~bohr.}
We considered a mesh of
  240 $k$-points in the irreducible Brillouin
zone (IBZ) corresponding to the space group $I4/mmm$ of the parent
compound unit cell.  Both $R_{\rm MT}K_{\rm max}$ and the number of
$k$-points in the IBZ were tested to be sufficient for rendering an
accurate electronic bandstructure.

The supercell has the symmetry of  a  centered monoclinic unit cell in the
space group $Cm$  which includes as symmetry operations   the
identity  and a mirror plane reflection
perpendicular to the atomic layers passing through the
interstitial oxygen atom.  The atomic displacements caused by the
interstitial oxygen are mirror symmetrical about this plane, as
illustrated in Fig.~\ref{local_hop} for Cu atoms.
 For the supercell calculations, we used the
same $R_{\rm MT}K_{\rm max}$ and $R_{\rm MT}$ values as for the parent compound and 64 $k$-points in the IBZ of the
supercell.

\section{Details of the tight-binding models and application to undoped {\BSCCO}}

The parent compound contains four Cu atoms per unit cell: two Cu
atoms per slab.  Since the {\BSCCO} lattice is base-centered and each
eigenvalue in the electronic structure is doubly degenerate due to the
presence of  two equivalent slabs, there are  only two  doubly degenerate Cu
$3d_{x^2-y^2}$ bands. We therefore  construct the TB Hamiltonian by
considering only one pair of Cu atoms in one slab:
\begin{equation}
 H=\sum_{{\bf k}}\left[
\begin{array}{cc}
 d_1^{\dagger}({\bf k})& d_2^{\dagger}({\bf k})
\end{array}\right]
\left[\begin{array}{cc}
       E_{xy}({\bf k})&E_{\bot}({\bf k})\\
       E_{\bot}(-{\bf k})&E_{xy}({\bf k})
     \end{array}\right]\left[
\begin{array}{c}
 d_1({\bf k})\\d_2({\bf k})
\end{array}\right],\label{Ham}
\end{equation}
where $d^{\dagger}_i({\bf k})$ [$d_i({\bf k})$] create
[annihilate] an electron with a wave-vector ${\bf k}$ in the Cu
$3d_{x^2-y^2}$ orbital of atom $i=1,2$ (Cu atoms 1 and 2 belong to
different CuO$_2$ layers). The energy dispersion
$E_{xy}({\bf k})$ is due to interactions between Cu atoms
within a layer and $E_{\bot}({\bf k})$ is due to inter-layer
electron hoppings; both dispersions are given by
\[
 E_{xy,\bot}({\bf k})=\sum_{{\bf l}}\exp(i{\bf k} \cdot {\bf l})t_{{\bf l}},
\]
where $t_{{\bf l}}$ is the hopping integral between Cu atoms that are
connected by a vector ${\bf l}$. This
vector  is denoted as ${\bf l}=(n,m,z)$ and corresponds
to a vector $(na,ma,zc)$ in absolute coordinates where $a$ and $c$
are the unit cell parameters, $n, m$  denote
integers and $z=0.099$,  is the
distance between  two CuO$_2$ layers in units of $c$.

By optimizing the values of the hopping integrals $t_{{\bf l}}$, the
eigenvalues of the Hamiltonian Eq.~(\ref{Ham}) can be adjusted to describe
the Cu $3d_{x^2-y^2}$ bands of the parent compound. In order to obtain
a description that satisfies our accuracy requirements, we had to
include into the model effective hopping integrals between 12
Cu nearest
neighbors as  listed in
Table~\ref{parent_hoppings1}. We denote this tight-binding results
TB1$_{\rm undoped}$.

The in-plane nearest-neighbor hopping integral $t_{100}=-0.5196$ eV
and second nearest-neighbor hopping integral $t_{110}=0.1115$ eV are
in agreement with the results of  previous DFT
calculations~\cite{other_bands} and with the analysis of photoemission
measurements of the {\BSCCO} electronic structure~\cite{Radtke_94}.
In Fig.~\ref{bands_f1}~(a), the energy spectrum of the TB
Hamiltonian is compared with the DFT bands. The agreement is overall
good, except for some small deviations around the $\Gamma$
point. This region  is particularly difficult to
reproduce since here the shape of the Cu $3d_{x^2-y^2}$ bands is
strongly influenced by the hybridizations between the Cu $3d_{x^2-y^2}$
orbital and other energetically close  Cu $d$  and O $p$
orbitals.

\begin{table}
\caption{TB1$_{\rm undoped}$ results: Optimized values of the on-site energy $\mu$ and the
hopping integrals $t_{{\bf l}}$ between 12 Cu nearest neighbors in eV.
These parameters reproduce the Cu $3d_{x^2-y^2}$ bands
in Fig.~\ref{bands_f1}~(a) and (b). The vector ${\bf l}=(n,m,z)$
is given by integers $n, m$; $z$ can take values of
$0$ or $z=0.099$ as $0.099c$ is the distance between  two CuO$_2$
layers.} \label{parent_hoppings1}

\renewcommand{\arraystretch}{1.4}
\begin{tabular*}{\columnwidth}{@{\extracolsep{\fill}}cccccc}
\hline \hline
$\mu$&$t_{00z}$&$t_{100}$&$t_{10z}$&$t_{110}$&$t_{11z}$\\\hline
0.4212& 0.0543& -0.5196& 0.0056& 0.1115& -0.0221 \\
\hline \hline
\end{tabular*}
\begin{tabular*}{\columnwidth}{@{\extracolsep{\fill}}ccccccc}
$t_{200}$&$t_{20z}$&$t_{210}$&$t_{21z}$&$t_{220}$&$t_{22z}$&$t_{300}$\\\hline
 -0.0859& 0.0117& -0.0078& -0.0064& 0.0025& -0.0103&-0.0238 \\
\hline \hline
\end{tabular*}
\end{table}

\begin{table}
\caption{TB2$_{\rm undoped}$ results: Optimized values of the on-site energy $\mu$ and hopping
  integrals $t_{{\bf l}}$  for the parent compound where only the closest nearest neighbors
are considered.  Also shown are the TB results obtained by mapping
the doped supercell bands to a homogeneous Hamiltonian (TB2$_{\rm hom.\;doped}$).
The meaning of the three subindices  is the same as in
Table~\ref{parent_hoppings1}.} \label{parent_hoppings2}
\renewcommand{\arraystretch}{1.4}
\begin{tabular*}{\columnwidth}{@{\extracolsep{\fill}}lccccc}
\hline \hline
&$\mu$&$t_{100}$&$t_{110}$&$t_{200}$&$t_{00z}$\\\hline
TB2$_{\rm undoped}$ & 0.4464& -0.5174& 0.1085& -0.0805& 0.0818 \\
TB2$_{\rm hom.\;doped}$ & 0.4900& -0.5150& 0.1158& -0.0800& 0.0700
\end{tabular*}
\begin{tabular*}{\columnwidth}{@{\extracolsep{\fill}}lccccc}
\hline \hline
&$t_{11z}$& $t_{210}$&$t_{300}$&$t_{400}$&$t_{21z}$\\\hline
TB2$_{\rm undoped}$ & -0.0264& -0.0073 & -0.0182& -0.0122& -0.0044 \\
TB2$_{\rm hom.\;doped}$ & -0.0229 &-0.0075 & -0.0177& -0.0046& -0.0062
\end{tabular*}
\begin{tabular*}{\columnwidth}{@{\extracolsep{\fill}}lcccc}
\hline \hline
&$t_{220}$&$t_{330}$&$t_{500}$&$t_{33z}$\\\hline
  TB2$_{\rm undoped}$ & 0.0068& -0.0052&-0.0049& -0.0047\\
 TB2$_{\rm hom.\;doped}$  & 0.0045& -0.0015& -0.0012& -0.0003 \\
\hline \hline
\end{tabular*}
\end{table}

\section{Details of the TB model for the O-doped supercell}

Here, we present the TB model for the O-doped supercell,
based upon the TB1$_{\rm undoped}$  model for the undoped single cell, whose
parameters we use as initial.  As adjustable
parameters, we choose three hopping integrals of the $t_{100}$ type,
two of the $t_{110}$ type and three of the $t_{200}$ type that connect
the Cu atoms experiencing the largest displacement due to the
interstitial oxygen and its neighbors (the hopping integrals of the
$t_{100}$, $t_{110}$ and $t_{200}$ types are represented by,
respectively, solid, dashed and dash-dotted lines in
Fig.~\ref{local_hop}). We also allow for different on-site energies $\mu$
for the $3d_{x^2-y^2}$ orbitals of the 8 Cu atoms in the CuO$_2$ layer
closest to the dopant.  Making use of the crystal symmetry, the number
of  $\mu$ parameters is reduced to six. We then  assign a unique $\mu$
value  to
the on-site energies of the other 8 Cu atoms since we expect that they
are less affected by the dopant.  These seven on-site energies
together with the eight hopping integrals are varied during the
Hamiltonian optimization.  The optimized values of the hopping
integrals are given in Fig.~\ref{local_hop} in eV and
the on-site energies are listed in the Figure caption. We denote this model TB1$_{\rm loc.\;doped}$.
 We observe  the largest  on-site
energy variation  for the most displaced Cu (marked with an arrow in Fig.~\ref{local_hop})
while Cu further away from the dopant are hardly affected.
The variations in the hopping integrals are also consistent.
 For example, $t_{100}$ increases when the two Cu
atoms get closer and  slightly decreases when they are pushed apart by the
interstitial oxygen.  The decrease is even stronger  when the Cu atoms
shift with respect to each other parallel to the mirror plane, what appreciably
reduces the orbital overlap.

 We also considered the same 15 parameters (7 $\mu$'s and 8 $t$'s)
 to map the DFT Cu $3d_{x^2-y^2}$ bands with a supercell TB
Hamiltonian based upon the TB2$_{\rm undoped}$ for the parent compound,
Table~\ref{parent_hoppings2}. While  a mapping to the
doped bands  is almost as good as the one given by the previous
model, the resulting model
parameters assumed seemingly chaotic values not consistent with their
expected behavior. One faces similar inconsistencies also when other
trial sets of adjustable parameters are used. The failure of the
TB2$_{\rm undoped}$ (Table~\ref{parent_hoppings2}) in
describing the dopant-induced changes in the bandstructure of
{\dopedBSCCO} indicates that the results provided by the approach
based on optimizing certain selected model parameters depend very
strongly on the
choice of effective far neighbor interactions that are not optimized.

\end{document}